\documentclass{aa}
\usepackage{natbib,psfig,graphicx}
\begin{document}

% \thesaurus{02.07.1; 03.13.8; 11.03.4;
% 11.05.1; 11.05.2; 11.06.1; 11.06.2; 11.19.6}

\title{The Vimos VLT Deep Survey: Compact structures in the CDFS}

\offprints{C. Adami \email{christophe.adami@oamp.fr}}

\author{
C. Adami\inst{1} 
\and A. Mazure\inst{1}  
\and O. Ilbert \inst{1,3}
\and A. Cappi  \inst{3} \\
%BUILDERS(b):
\and D. Bottini \inst{2}
\and B. Garilli \inst{2}
\and V. Le Brun \inst{1}
\and O. Le F\`evre \inst{1}
\and D. Maccagni \inst{2}
\and J.P. Picat \inst{7}
\and R. Scaramella \inst{4}
\and M. Scodeggio \inst{2}
\and L. Tresse \inst{1}
\and G. Vettolani \inst{4}
\and A. Zanichelli \inst{4}\\
%SURVEY CORE(s):
\and M. Arnaboldi \inst{5}
\and S. Arnouts \inst{1}
\and S. Bardelli  \inst{3}
\and M. Bolzonella  \inst{6} 
\and S. Charlot \inst{8,10}
\and P. Ciliegi    \inst{3}  
\and T. Contini \inst{7}
\and G. Covone \inst{1}
\and S. Foucaud \inst{2}
\and P. Franzetti \inst{2}
\and I. Gavignaud \inst{7,12}
\and L. Guzzo \inst{9}
\and A. Iovino \inst{9}
\and S. Lauger \inst{1}
\and H.J. McCracken \inst{10,11}
\and B. Marano     \inst{6}  
\and C. Marinoni \inst{1}
\and B. Meneux \inst{1}
\and R. Merighi   \inst{3} 
\and S. Paltani \inst{1}
\and R. Pell\`o \inst{7}
\and A. Pollo \inst{9}
\and L. Pozzetti    \inst{3} 
\and M. Radovich \inst{5}
\and G. Zamorani \inst{3} 
\and E. Zucca    \inst{3}\\
%ASSOCIATES(a):
\and M. Bondi \inst{4}
\and A. Bongiorno \inst{6}
\and G. Busarello \inst{5}
\and L. Gregorini \inst{4}
\and G. Mathez \inst{7}
\and Y. Mellier \inst{10,11}
\and P. Merluzzi \inst{5}
\and V. Ripepi \inst{5}
\and D. Rizzo \inst{7}
}

\institute{
1) Laboratoire d'Astropysique de Marseile, UMR 6110 CNRS-Universit\'e de
Provence,  BP8, 13376 Marseille Cedex 12, France
\and
2) IASF-INAF - via Bassini 15, I-20133, Milano, Italy
\and
3) INAF-Osservatorio Astronomico di Bologna - Via Ranzani,1, I-40127, Bologna, Italy
\and
4) IRA-INAF - Via Gobetti,101, I-40129, Bologna, Italy
\and
5) INAF-Osservatorio Astronomico di Capodimonte - Via Moiariello 16, I-80131, Napoli,
Italy
\and
6) Universit\`a di Bologna, Dipartimento di Astronomia - Via Ranzani,1,
I-40127, Bologna, Italy
\and
7) Laboratoire d'Astrophysique de l'Observatoire Midi-Pyr\'en\'ees (UMR 
5572) -
14, avenue E. Belin, F31400 Toulouse, France
\and
8) Max Planck Institut fur Astrophysik, 85741, Garching, Germany
\and
9) INAF-Osservatorio Astronomico di Brera - Via Brera 28, Milan,
Italy
\and
10) Institut d'Astrophysique de Paris, UMR 7095, 98 bis Bvd Arago, 75014
Paris, France
\and
11) Observatoire de Paris, LERMA, 61 Avenue de l'Observatoire, 75014 Paris, 
France
\and
12) European Southern Observatory, Karl-Schwarzschild-Strasse 2, D-85748
Garching bei München, Germany
}

\date{Accepted . Received ; Draft printed: \today}

\authorrunning{Adami et al.}
\titlerunning{Compact structures in the CDFS: dynamical state and content}

\abstract{We have used the Vimos VLT Deep Survey in combination with other spectroscopic,
  photometric and X-ray surveys from literature to
detect several galaxy structures in the Chandra Deep Field South (CDFS).
Both a friend-of-friend based algorithm applied to the
spectroscopic redshift catalog and an adaptative kernel galaxy density and colour
maps correlated with photometric redshift estimates have been used. 

We mainly detect a chain-like structure at z=0.66 and two massive groups at 
z=0.735 and 1.098 showing signs of ongoing collapse.
We also detect two galaxy walls at z=0.66 and at z=0.735 (extremely compact in
redshift space). The first one contains the chain-like structure and the last 
one contains in its center one of the two massive groups. 
Finally, other galaxy structures that are probably loose low mass groups are detected. 

We compare
the group galaxy population with simulations in order to assess the richness
of these structures and we study their galaxy morphological contents. The
higher redshift structures appear to probably have lower velocity dispersion 
than the nearby ones. The number of moderatly massive structures we detect is
consistent with what is expected for an LCDM model, but
a larger sample is required to put significant cosmological constraints.}

\maketitle


\section{Introduction}

Modern galaxy surveys usually map the large scale structure of the
Universe over large contiguous regions at low redshifts (e.g. Folkes et
al. 1999 for the 2dF survey or Castander 1998 for the SDSS) or more
sparsely up to z$\sim$1.
These surveys use for example clusters of galaxies (e.g. Romer et al. 2001)
or larger scale filaments and walls 
(e.g. Dav\'e et al. 1997) to constrain cosmological models. However, it is very rare to be able to
combine multi-wavelength data over relatively large contiguous regions
in order to
search homogeneously for
structures up to z$\sim$1. Similarly, spectroscopic
samples are usually very incomplete inducing several detection biases.

In this framework, the CDFS area has become these last years an 
intensively surveyed area in several wavelengths: from
X-rays (e.g. Giacconi et al. 2002) to optical and near infrared
(e.g. Moy et al. 2003, Arnouts et al. 2001), both in imaging and  
spectroscopic modes (e.g. Gilli et al. 2003, Le F\`evre et al. 2004). 
Recently, a very large catalog of 1599 spectra was released by the VVDS team (Le
F\`evre et al. 2004). These new redshifts were measured in an area of
21'$\times$21.6' and include a total of 1452 galaxies, 139 stars and 8
QSOs. The redshift distribution is peaked at a median redshift of 0.73 and
include measurement down I$_{AB}$=24. Combination of these data makes now
possible galaxy structure identifications in this field with a five
times larger spectroscopic data sample as the one used for example in Gilli et 
al. (2003). 
In a companion paper (Scaramella et al. 2006) very large scale and diffuse 
galaxy structure will be presented while we concentrated on compact structures
(groups, clusters and compact walls) in this paper. 

Detecting groups or clusters is not, however, an easy task, essentially
because these are not well defined objects but assemblies of objects. Even if
they can best be defined as "potential wells", which are responsible for
lensing effect and X-ray emission of the hot gas, the use of these techniques 
requires, especially for distant systems, complementary photometric and spectroscopic
observations of the candidate member galaxies in order to measure their redshift and
colors.

Historically (e.g. Abell 1989 and references therein), clusters or groups
appear as excess in the galaxy density field but the use of the "number
density excess" with respect to the background as a method of detection becomes
rapidly inefficient as soon as more distant systems are searched. Indeed, the contrast
is decreasing as long as the apparent limiting magnitude is increasing. This
difficulty is still present when using the third dimension (in terms of
redshift) since, even if virialized structures extend on more or less six times
their velocity dispersion in redshift space, any survey probes less and less
clusters member galaxies in apparent magnitude limited surveys.  

We present our results based on public spectroscopic redshifts,
Combo17 photometric data (e.g. Wolf et al. 2004) and X-ray
source catalogs (Giacconi et al. 2002). In this paper, we combine these data
with our own spectroscopic
observations of the CDFS field (Le F\`evre et al. 2004) available at the
CENCOS database (http://cencosw.oamp.fr/FR/index.fr.html). The methods
developed in this paper will be used to search and study clusters in larger
samples such as the complete Virmos VLT Deep Survey (VVDS hereafter).

Section 2 presents the detection methods and the samples we used. Section 3
presents the structure analysis methods. Sections 4
reviews the galaxy structures we detected. The final discussion of our results
is in Section 5. All quantities were calculated assuming a standard flat LCDM
model with H$_0$=65 km/s/Mpc, $\Omega _m$=0.3, $\Omega _{\Lambda}$=0.7.

\begin{figure}
\centering
%\mbox{\psfig{figure=compcdfsclean.ps,width=8cm,angle=270}}
\caption[]{Galaxy redshift sampling rate for
I magnitudes between 17 and 24 in the CDFS. Completeness is given in percentage on
the vertical axis.}
\label{fig:comp}
\end{figure}

\begin{figure}
\mbox{\psfig{figure=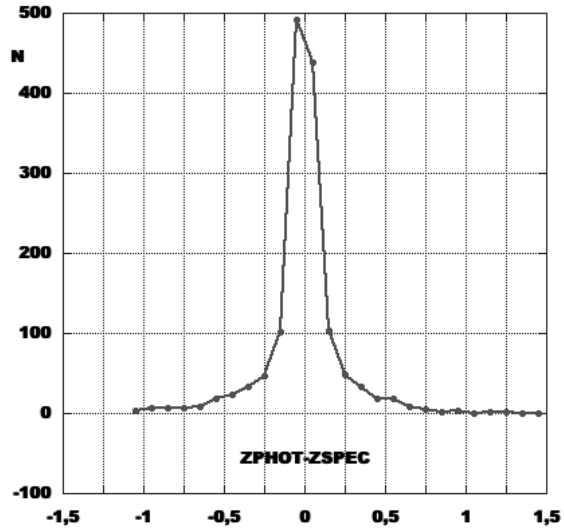,width=8cm,angle=0}}
\caption[]{Histogram of the difference between spectroscopic and photometric
  redshifts in the CDFS using Combo17 magnitudes down to I=24.}
\label{fig:specphot}
\end{figure}

\begin{figure}
\caption[]{Upper figure: 3D view of the spectroscopic catalog distribution. 
Lower figure: 3D view of the photometric redshift catalog
  distribution. The thin z=0.73 wall is visible in the upper figure and 
merged with the z=0.66 large structure in the lower figure.}
\label{fig:CDFS3D1D}
\end{figure}

%\begin{figure}
%\centering
%\mbox{\psfig{figure=fig_em.ps,width=10.5cm,angle=270}}
%\mbox{\psfig{figure=fig_abs.ps,width=10.5cm,angle=270}}
%\caption[]{Percentage of emission line only (upper figure) and
%absorption line only (lower figure) galaxies as a function of
%redshift. Solid line: field galaxies. Open (less than 10 galaxies in
%the structure) and solid circles (more than 10 galaxies in
%the structure): detected structures. Dashed vertical lines show in the upper
%figure the redshift where main espectral lines are exiting our spectral range. Dashed
%vertical lines show the redshift where main spectral lines are entering our 
%spectral range.} 
%\label{fig:absem}
%\end{figure}

\begin{figure}
\centering
%\mbox{\psfig{figure=gap.ps,width=8cm,angle=0}}
\caption[]{Gap redshift histogram in the CDFS spectroscopic survey.}
\label{fig:gap}
\end{figure}

\begin{figure}
\centering
%\mbox{\psfig{figure=S1morph.ps,width=7cm,height=3.5cm,angle=270}}
\mbox{\psfig{figure=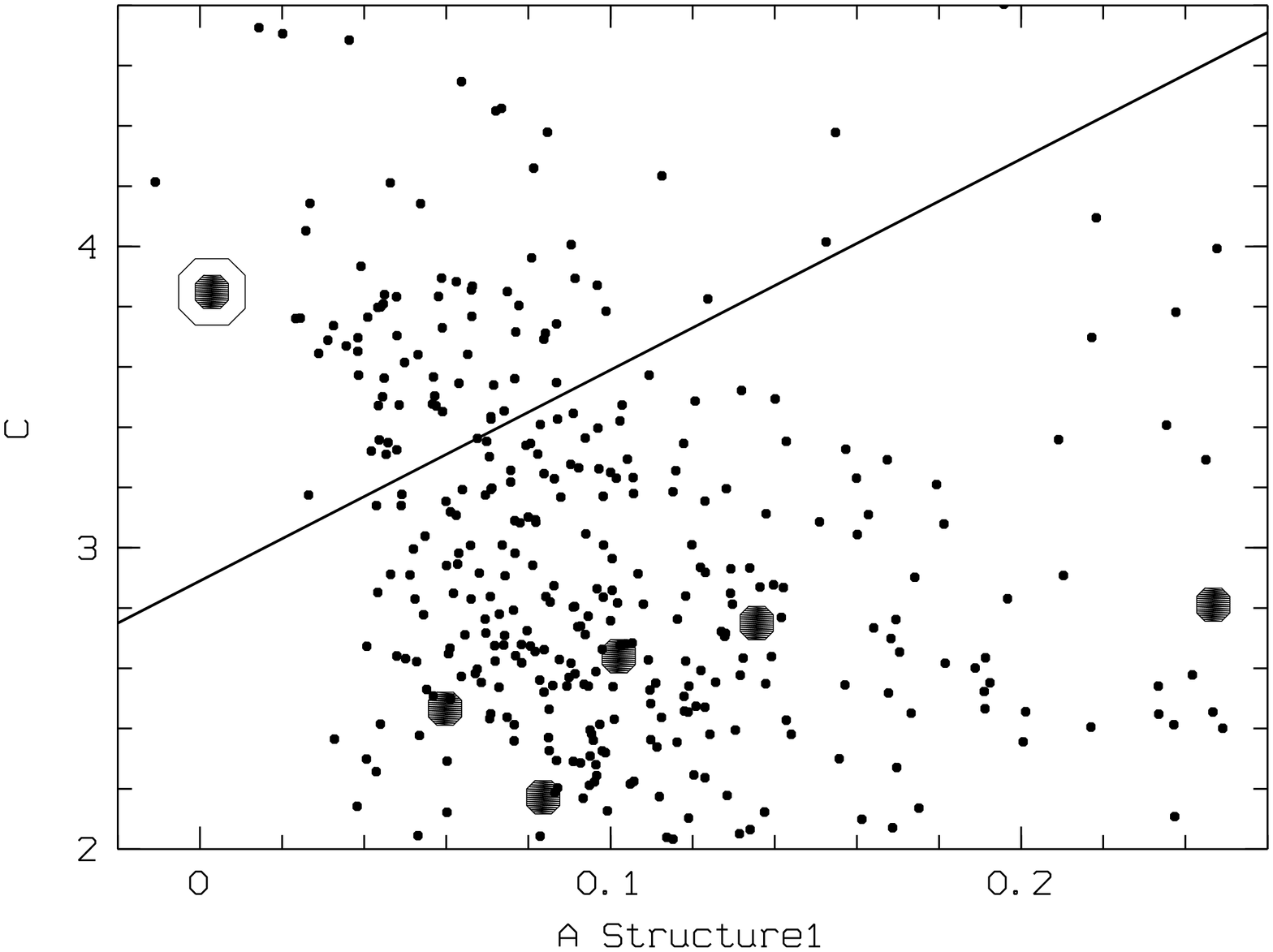,width=9cm,angle=270}}
%\mbox{\psfig{figure=S6morph.ps,width=7cm,height=3.5cm,angle=270}}
\mbox{\psfig{figure=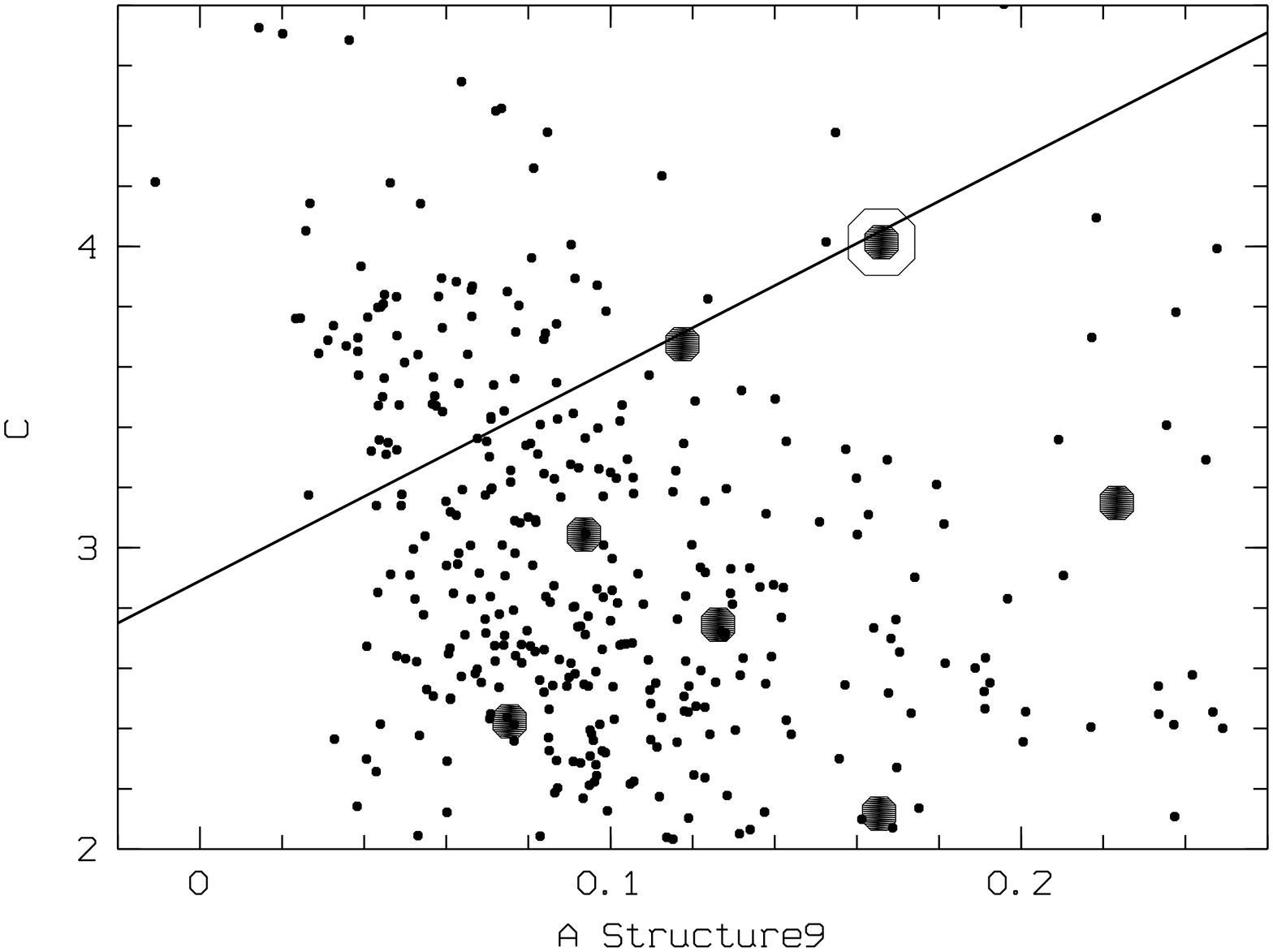,width=9cm,angle=270}}
%\mbox{\psfig{figure=S10morph.ps,width=7cm,height=3.5cm,angle=270}}
\mbox{\psfig{figure=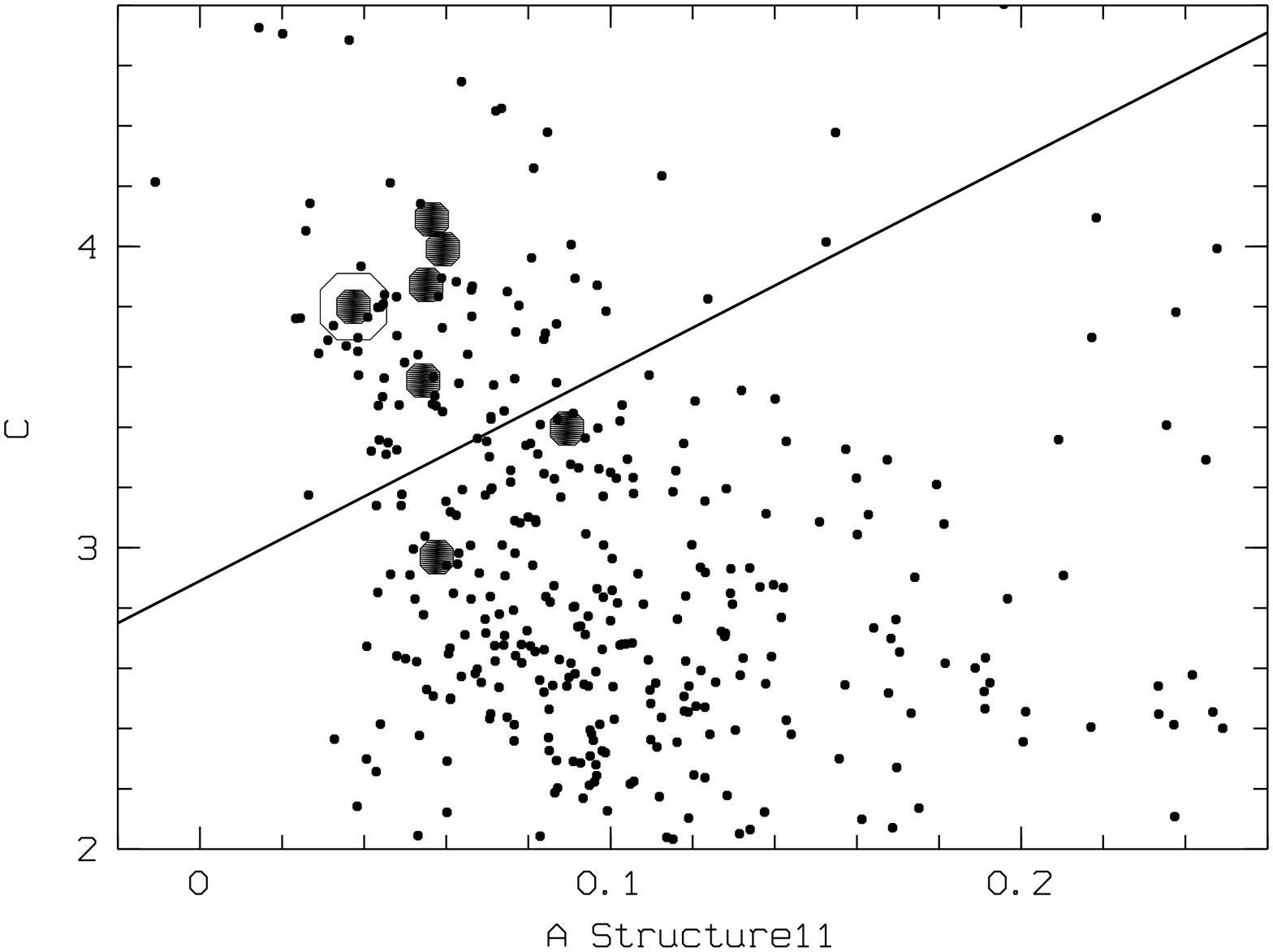,width=9cm,angle=270}}
%\mbox{\psfig{figure=Swallmorph.ps,width=7cm,height=3.5cm,angle=270}}
\caption[]{Asymmetry versus Concentration of galaxies in Structures 1, 9 and 11. The solid line
symbolizes the separation between early and late morphological
types. Filled circles are the galaxies inside the considered
structure, circled filled circle is the brightest galaxy of the
structure and dots are all the field galaxies 
along the CDFS line of sight.} 
\label{fig:morph}
\end{figure}

%\begin{figure}
%\centerline{\hbox{\psfig{file=cdfs2.ps,width=5.5cm,angle=270}
%\psfig{file=histo_s2.ps,width=3.5cm,angle=270}}}
%\caption[]{Color Magnitude Relation (CMR) for 
%Structure 1. X-axis is the B magnitude and y-axis is the B-V
%color. Dots are field galaxies with measured spectroscopic
%redshift. Filled circles are galaxies within the considered
%structure. Right panels are the redshift histograms with a step of 300
%km/s}
%\label{fig:cdfslow}
%\end{figure}

\begin{figure}
\mbox{\psfig{figure=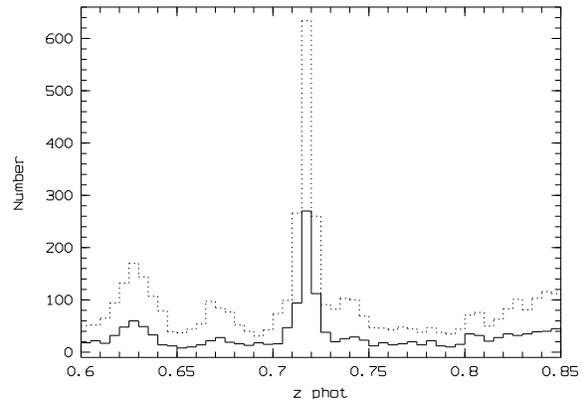,width=8cm,angle=270}}
\caption[]{Histograms of photometric redshifts around the z=0.73
  wall. Dashed line: inside the spectroscopic area, solid line: outside the
  spectroscopic area and inside the photometric redshift area.}
\label{fig:histoinout}
\end{figure}

\begin{figure}
\mbox{\psfig{figure=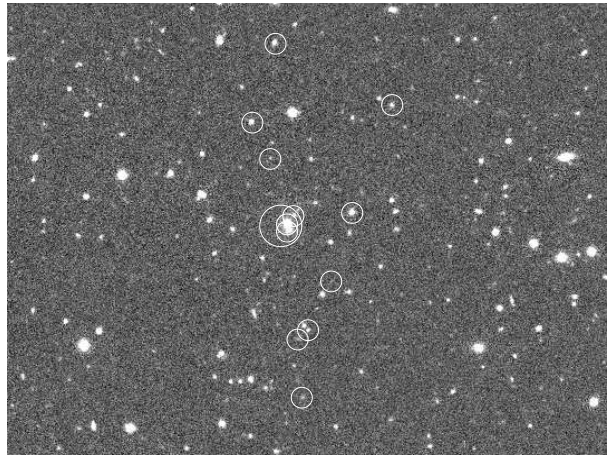,width=8cm,angle=0}}
\caption[]{HST I image of Structure 11-1 at z=0.73 with galaxies associated to the
structure (small circles) and the X-ray source (large circle). Size of the
image is 5.1'$\times$3.4'.}
\label{fig:11-1fig}
\end{figure}

\begin{figure}
%\centerline{\hbox{\psfig{file=cdfs9.ps,width=12cm,angle=270}
%\psfig{file=histo_s9.ps,width=8cm,angle=270}}}
\psfig{file=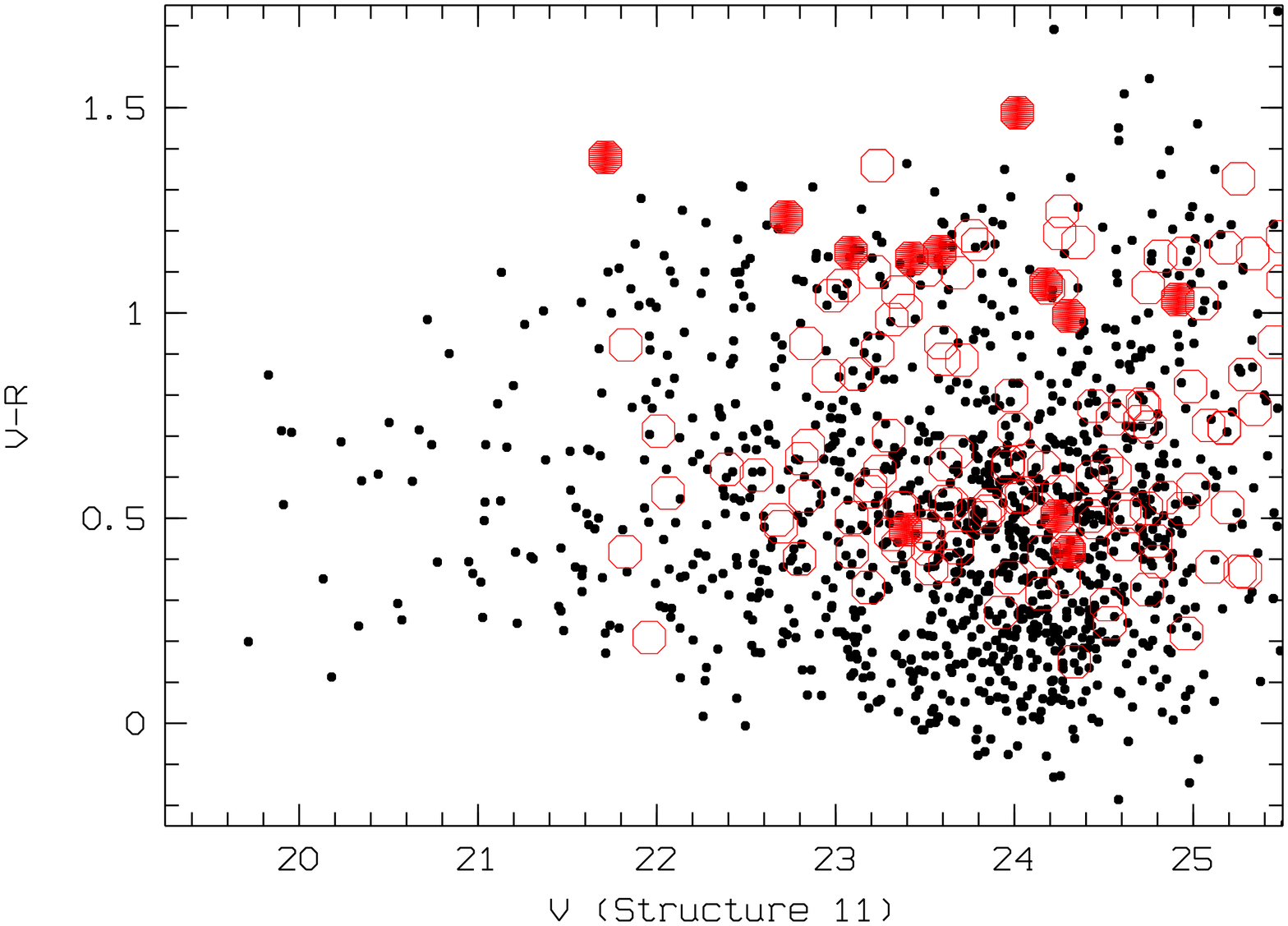,width=9cm,angle=270}
\psfig{file=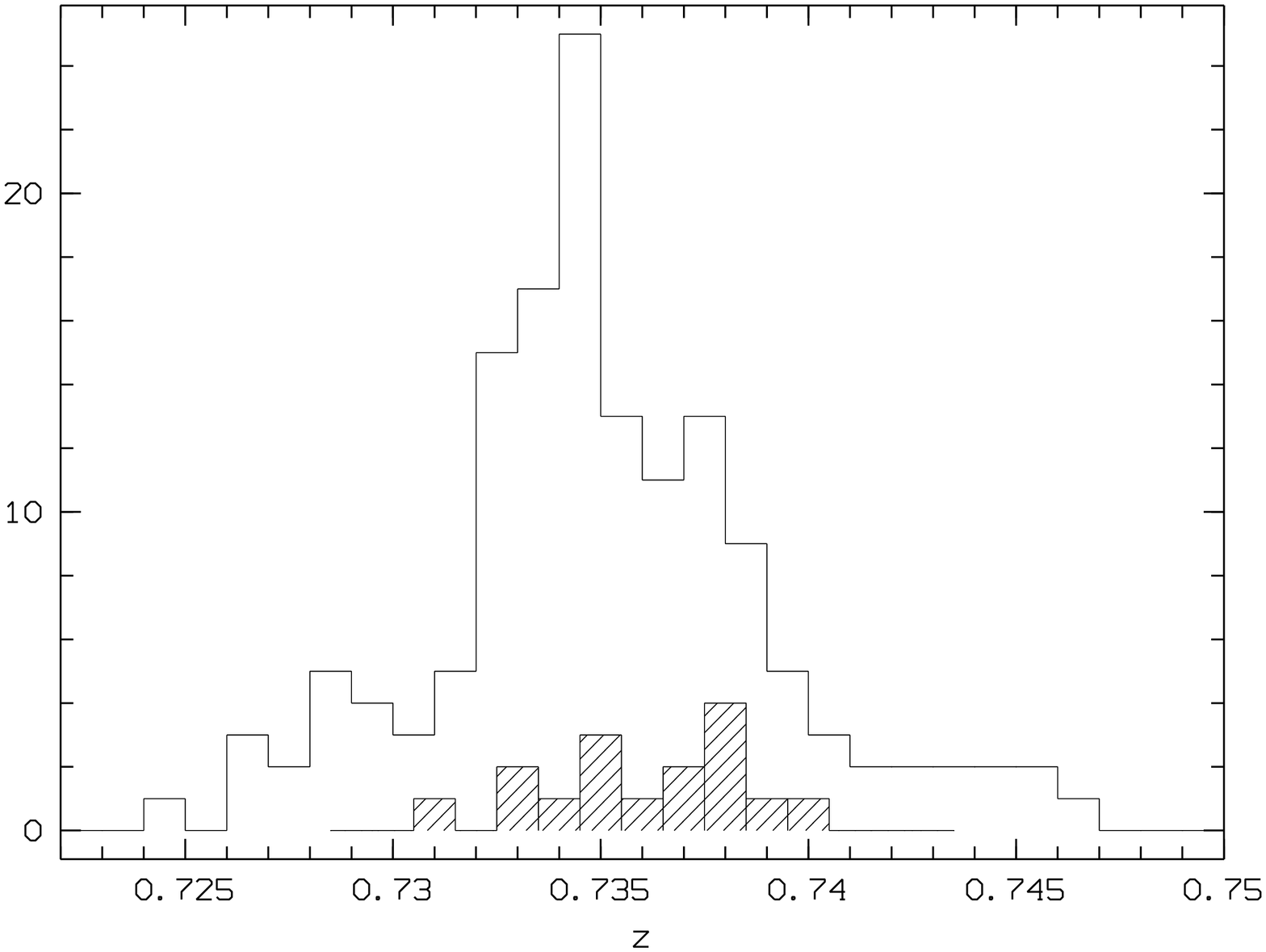,width=9cm,angle=270}
\caption[]{Top: color Magnitude Relation (CMR) for 
Structure 11. X-axes are the V magnitudes and y-axes are the V-R
colors. Dots are field galaxies with measured spectroscopic
redshift. Structure 11 (the z=0.735 wall) is
shown as open circles while the central group (Structure 11-1) is
shown as filled circles. Bottom: redshift histogram with a step of 300
km/s (dashed histogram is the structure 11-1). 
%Bottom: B-R color isodensity
%map. Detected structures in the z=0.735 wall are shown as filled circles.
}
\label{fig:cdfsinter}
\end{figure}

\begin{figure}
\psfig{file=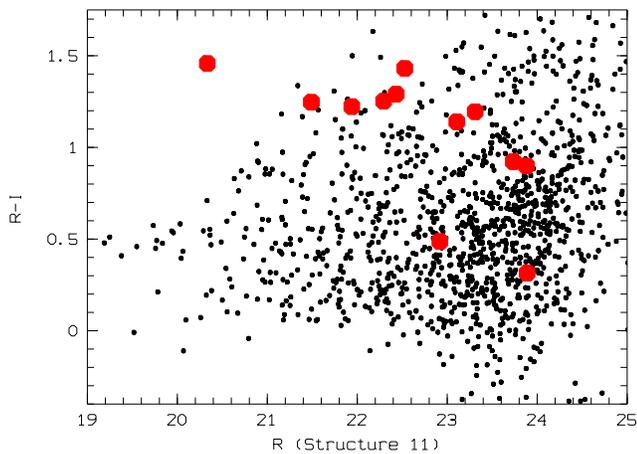,width=9cm,angle=270}
%\psfig{file=histo_s11.ps,width=8cm,angle=270}
%\centerline{\hbox{\psfig{file=cdfs15.ps,width=12cm,angle=270}
%\psfig{file=histo_s15.ps,width=8cm,angle=270}}}
\caption[]{Same as Fig.~\ref{fig:cdfsinter} with R/R-I CMR and for Structure 11.}
\label{fig:cdfshigh}
\end{figure}

\begin{figure}
\centering
\mbox{\psfig{figure=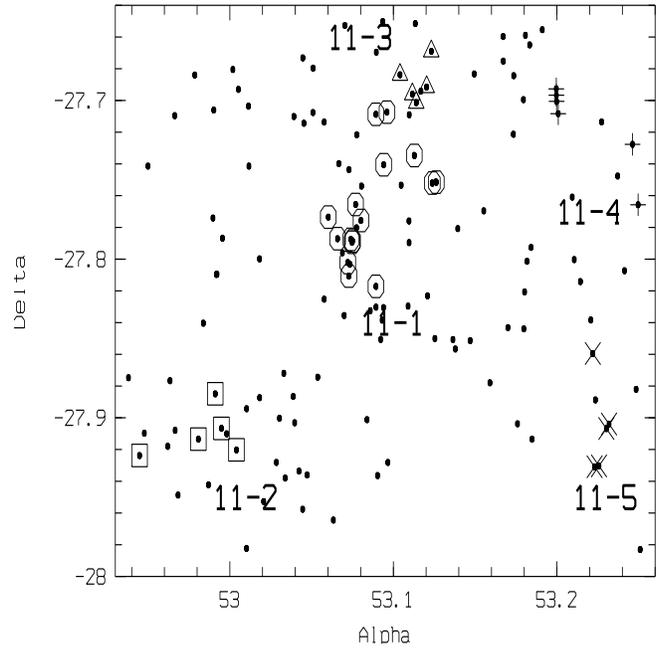,width=9.cm,height=9cm,angle=270}}
\caption[]{Dots: from spectroscopy, all galaxies inside the z$\sim$0.735 wall (Structure
11). Circled dots: main dynamical group. Squares, triangles and
crosses: additionnal dynamically linked groups.} 
\label{fig:S11}
\end{figure}

\begin{figure}
\centering
\mbox{\psfig{figure=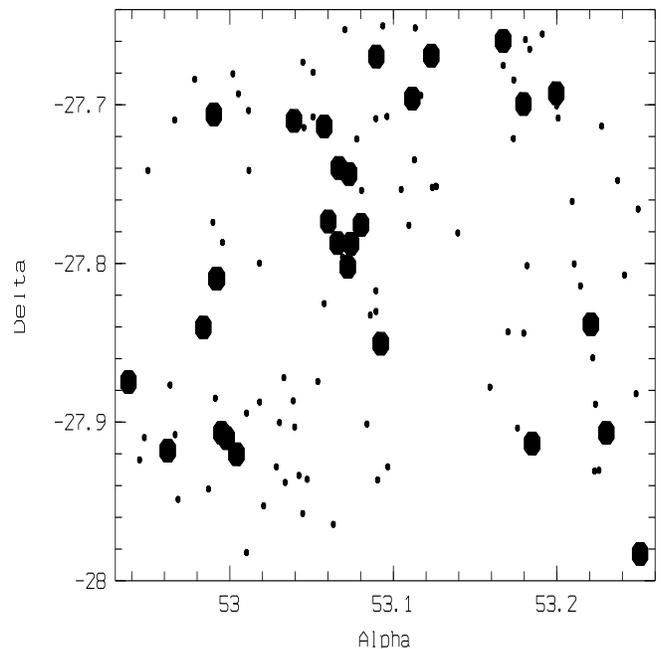,width=9.cm,height=9cm,angle=270}}
\caption[]{Dots: all galaxies inside the z$\sim$0.735 wall (Structure
11). Filled circles: early type galaxies from morphological classification.} 
\label{fig:S11bis}
\end{figure}

\begin{figure}
%\mbox{\psfig{figure=b073.ps,width=6cm,angle=0}}
%\mbox{\psfig{figure=r073.ps,width=6cm,angle=0}}
%\mbox{\psfig{figure=comp2D.ps,width=6cm,angle=0}}
\caption[]{Adaptative kernel density contours of the z=0.735 wall from
  photometric redshifts. East is to the right and north is to the top. Top: galaxies with 
B-R=[1.5;2] are overplotted. Middle: galaxies with B-R greater than 2
are overplotted. Bottom: 2D spectroscopic redshift sampling rate:
the darker the color code, the higher the sampling rate, the lighter the
color code, the lower the sampling rate. The sampling rate is typically
varying from 15 to 30$\%$.}
\label{fig:specphot2}
\end{figure}

\section{Detection methods and structure galaxy sample}

\subsection{What we are looking for: aims and strategy}

We present here several key numbers. A galaxy structure appears on the sky 
(counted in 2D) as an excess in the galaxy density field.
For a given limiting magnitude, this contrast
is decreasing with redshift. Table 1 shows for example this contrast for
a Coma-like cluster put at z=0.5 and 1 for several limiting apparent
magnitudes (assuming no evolution).

However, assuming that the red sequence (corresponding to early type objects) 
is present at least in the rich structures, its detection in photometric bands
matching the expected redshift either using color magnitude relation or color
density maps (as described below) is a potentially adequate method. 

Contrarily to photometric catalogs that provide almost uniform coverage on broad
regions of the sky, uniform spectroscopic coverage (and especially of dense
areas as groups or clusters) even with the most efficient instrumentation needs
a considerable effort to ensure a sufficient sampling both in terms of space
and magnitude range. Consequently, the characteristics of the detected
structures clearly suffer from the bias induced by observational
procedures. The VVDS aims to overcome these difficulties by
using an optimal strategy in terms of spatial coverage and redshift sampling
(e.g. Le F\`evre et al. 2004). The resulting effects on the detection of
structures have already been discussed in Rizzo et al. (2004).

\subsection{The spectroscopic sample}

In order to search for compact structures inside the Chandra Deep Field
South we used all the available redshifts in the literature in this region. 

Starting from a VVDS catalog of 1460 redshifts with quality flags greater than
1 (i.e. greater than 75$\%$ confidence level, see Le F\`evre et al. 2004) and
using an identification distance of 1'', we added 
129 additionnal redshifts from the ESO-GOODS sample using spectra with
quality strictly greater than 0.5, 222 redshifts from Szokoly et al. (2004)
measured on spectra with qualities equal to A and B and 7 redshifts from the 2dF
survey. In total, we are therefore using a catalog of 1818 
redshifts. This represents about 25$\%$ of all galaxies in
the CDFS field of view down to I=24. 
As demonstrated in Scaramella et al. (2006), these catalogs
are in very good agreement regarding redshift estimates 
and merging them is, therefore, not a concern. 

As a first step, we limited this catalog to redshift greater than
0.16 as it 
was very unlikely to detect interesting large scale features at lower
redshift due to our 
small angular extent ($\sim$18'$\times$19'). Moreover, there is an
Abell cluster (A3141) at 
z$\sim$0.11 only at 69' from the CDFS field center and the infalling
field galaxies onto this cluster are probaly significantly populating 
the CDFS area. The final catalog 
contains 1591 redshifts greater than 0.16.
The result of such a building process is that the
sampling rate of the galaxy redshift catalog is quite 
inhomogeneous across the field of view (Fig.~\ref{fig:comp}) because
all the surveys we used are not
covering the same area and have different sampling rates. VVDS data are quite
homogeneous over a large field while other surveys are uniform only over
smaller areas. However, as we are searching for distant structures, the more
galaxy redshifts we have, the more efficient is their detection.

As discussed above, even using redshifts, detection of distant structures (for
which m$^*$ is close to the survey limiting magnitude) is not
easy. Disentangling structure members from field galaxies and estimating the
velocity dispersion requires uncertainties on velocities to be well
estimated. Here we took advantage of having galaxies measured at least twice
to have such a realistic estimate. Using the 187 galaxies observed two times
and with a 75$\%$
confidence redshift estimate from both observations, it turns out that we have 
a dispersion close to 0.002, not strongly dependent on magnitude nor redshift
up to z$\sim$1.2. We adopted this value as the spectroscopic redshift uncertainty for
all galaxies. Uncertainties listed in Table 4 and 5 take into account this value.

  We note that the K20 data (e.g. Cimatti et al. 2002) were not public at
  the time of the present work completion. We plan however to use these data
  to search for very distant structures in the CDFS in a future article.

\subsection{The photometric samples}

We used the Combo-17 magnitude catalog (e.g. Wolf et al. 2004). This catalog
provides 17 photometric bands. We limited ourselves to I=24 which is the spectroscopic magnitude
limit. Using the Combo-17 photometry, photometric redshifts have been
calculated and compared to spectroscopic redshifts using the code "LePhare"
(http://www.lam.oamp.fr/arnouts/LE-PHARE.html). When limiting the samples
to z$_{spec}$=1.2, it turns out (see Fig~\ref{fig:specphot}) that in about
75$\%$ of cases, the difference is less than 0.15. Moreover, we used all
spectroscopic spectra with a confidence level greater than 75$\%$. This means
that we are not exempt from
spectroscopic errors in redshift estimates, increasing artificially this
difference of less than 0.15. 

Using the spectroscopic sample and the Combo17 magnitudes, we were also
able to compute a "photometric" type (see Section 3.3.2). Both photometric
redshift and photometric type estimates are distinct from the ones performed by
the Combo17 team and will be described in a future paper.

As the spatial coverage of photometric data is uniform, we take advantage of
the good quality of photometric redshifts to complement our spectroscopic
analysis by performing density and color maps within photometric redshift
slices (see next section). Structures detected with several methods
(spectroscopic AND photometric) have a better chance to be real. However, due
to uncertainties of photometric redshifts, real structures will be both
broadened and contaminated by adjacent photometric redshift slices. Advantages
(sharpness) and disadvantages (sparse sampling) of spectroscopic redshifts
versus advantages (good sampling) and disadvantages (broadening) of
photometric redshifts are illustrated in Fig~\ref{fig:CDFS3D1D} where the 3D
distribution of galaxies is plotted
using spectroscopic and photometric redshifts. In both cases, the main large
scale structures are visible (see following discussions).

\subsection{The spectroscopic structure detection method}

The method we used to detect stuctures is a friend of
friend based algorithm described in Adami $\&$ Mazure (2002)
and Rizzo et al. (2004). This algorithm first makes a
classical count-in-cells of
the galaxies in the redshift catalog (using a tunable window size on the sky and
in the redshift space). The cell size projected on the sky is chosen according to the type of
structures searched for (for clusters of galaxies, it corresponds to the 
characteristic size of such structures at a given redshift: we adopted a value
of 1Mpc). The size in
redshift space is also characteristic of the structures but can also be
infered from the redshift difference between the closest galaxy pairs 
in the sample (see
Fig~\ref{fig:gap}). The peak of galaxies below a given redshift separation
gives a statistical definition of the typical size of structures.
For the present case, we used 0.0026 (the value after
which the distribution in Fig~\ref{fig:gap} becomes more or less constant). We note that this
value is probably affected by the non-homogeneous redshift sampling rate
across the field of view.
If the number of galaxies in a given cell is large enough,
this cell is kept as significantly populated. Then, a percolation algorithm is
associating individual adjacent cells in larger structures. 
We chose in our analysis parameters adapted to the detection of
compact structures. The parameters used are similar to the ones given in 
Rizzo et al. (2004). The maximum redshift extension of a structure was
  fixed to $\Delta z = 0.02$. 
Elementary structures were merged when closer than
  2Mpc. The minimum number of galaxies per elementary structure was set to 4
  and the minimum number of galaxies per percolated structure (final
  structure) was set to 5. The size of the individual cells (where elementary
  structures were searched) was fixed to 2 Mpc, which was converted to
 the corresponding angular size as a function of redshift; the search was 
done in
different redshift slices: [0.16,0.31],  [0.31,0.40], then between z=0.40
and z=1.70 in slices of width $\Delta z =0.10$. We also checked separations
between each slices, detecting for example in this way S15 (z=1.098).

The physical parameter that will limit the
completeness of the structure catalog is the completeness of the redshift
catalog. For a constant galaxy sampling in terms of redshift
measures, we will for example more easily detect the richest structures
because these will be sampled with more redshifts. The
velocity dispersion is not by itself, however, a limiting parameter (besides
the fact that poor structures have generally low velocity
dispersions). We fixed the minimal number of galaxies with a redshift inside a
structure to 5 (i.e. structures with 4 galaxies or less are
automatically excluded as considered as spurious detections). 

We do not attempt to precisely estimate the structure detection rate in
this paper because of the complex galaxy redshift catalog sampling
rate. However,
following Rizzo et al. (2004), the structure detection rate should be lower
than 70$\%$ at z=0.3, lower than 50$\%$ at z=0.6 and about 10$\%$ for
z greater than 1. We remark,
as well, that nearly all structures detected with this method are
probably real with only about 10$\%$ expected to
be spurious (Rizzo et al. 2004), independently of the structure
apparent richness.

We note that a by-product of this structure detection procedure is a
field galaxy redshift catalog, i.e. the catalog of galaxies not included in any
compact structures.

\subsection{The photometric structure detection method}

While the previous method allows to disantangle close structures along the
line of sight thanks to the precision of the spectroscopic redshifts, this
method is limited by the completeness level of the spectroscopic sample. Low
sampling regions will not provide, therefore, a good structure detection
rate. In order to solve this problem, we used a second method based on
photometric redshift estimate and adaptative kernel color and galaxy density maps.

Adaptative kernel galaxy density maps are a common way to detect and study
nearby galaxy structures (e.g. Adami et al. 1998(a)). We estimated a
significance level for these maps using a bootstrap technique (e.g. Biviano et
al. 1996) with 1000 resamplings. This method is very
efficient as long as the detectable cluster galaxy population is dominating the
detectable field galaxy population. When trying to detect distant structures, however,
the cluster/field galaxy ratio becomes very low and structures
becomes virtually undetectable. In order to increase this ratio, we used the
photometric redshift estimates. We proceed in two steps: 

-First, we divide the
galaxy catalog into slices of $\pm$0.1. Such a width is similar to the
depth of nearby galaxy catalogs used to search for nearby structures
(e.g. Adami et al. 1998(a)) and is larger than the mean photometric redshift
uncertainty for individual galaxies.

-Second, we use the classical adaptative kernel galaxy density map method to
 eyeball the galaxy overdensities in the given redshift bin.

This method is applied to a complete photometric redshift catalog, so we are not 
affected by
inhomogeneous sampling rate. However, our photometric redshift estimate is
less precise than the spectroscopic estimate (even if one of the most precise
ever computed: see Fig~\ref{fig:specphot}),
and this, therefore, induces
a smoothing of the redshift distribution and reduces the ability to
disantangle close structures along the line of sight. We show in 
Fig.~\ref{fig:CDFS3D1D} the distribution along the
CDFS line of sight of the spectroscopic and photometric redshifts to
illustrate this problem. Similar structures are visible in both catalogs, but
photometric redshift distribution is less precise. The z=0.66 and z=0.735 walls
are for example almost merged in a single structure using photometric redshifts only,
while we create fake concentrations around z=0.87.

\subsection{Color spatial distribution}

As discussed above, rich clusters are generally characterized by an
excess of red objects even to z$\sim$1. The colors of elliptical galaxies
are given in Table 2 for
several redshifts (magnitude bands are chosen to encompass the 4000$\AA$ break
given the redshift). As a generalisation of the previous method, we also
used the mean color maps in
the $\pm$0.1 photometric redshift slices using the color corresponding to
the mean redshift. Computed with the adaptative kernel
method, this allows us to locate immediately the early type galaxy
concentrations in the redshift slices (assuming that red objects are
preferentially early type objects) and to compare with simple galaxy
overdensities. It also allows us to compare candidate structure locations with the
location of the structures detected with the spectroscopic method.

\section{Structure analysis methods}

\subsection{Galaxy velocity dispersion and spatial extension of the
detected structures}

For each of the detected structures, we computed the mean redshift, the
cosmological velocity dispersion (with Biweight estimators
from Beers 
et al. 1990) and the uncertainty on this value (using ROSTAT package
with 1000 bootstraps, Beers 
et al. 1990). This uncertainty is a 1-$\sigma$ error. We also computed the
spatial 
extension (computed as the maximum of the alpha and delta 
distribution second momentum) and the mean coordinates. All these
values are given in 
Tables 4 and 5. We note that even using robust estimators, we gave
confidence 
to the velocity dispersion estimate only when using more than 10
redshifts (e.g. Lax 1985).

\subsection{Red sequences in Color Magnitude Relations (CMR)}

Since the works of Baum (1959) and Sandage (1972) showing a correlation
between galaxy type, magnitude and color for early type galaxies in clusters, the CMR
is commonly used to characterize the cluster galaxy
populations. Relatively old and virialized clusters have an
old elliptical population and, therefore, exhibit a red sequence in the CMR
(at least at low and moderate redshifts). 

We apply the same test here with different magnitudes. Low redshift
structures will be studied using the B and V magnitudes (Wolf
et al. 2004), intermediate redshift
structures with the V and R magnitudes and high redshift
structures with the R and I magnitudes. These filter-pairs are
chosen in order to encompass the 4000$\AA$ break according to the
structure redshift. We used the Combo17 magnitudes because this is the
only homogeneous and unbiased photometric sample from B to I in our
whole field of view. Examples of expected values are given in Table 2. 

We note that a large part of the structure sample has
not a very early galaxy content. This makes impossible to define
properly a red sequence and, therefore, to compute homegeneously for the whole
sample a slope and a compactness. 
However, a possible test is to compare, using a 2D Kolmogorov
Smirnov test, the structure galaxy  
distribution in the magnitude/color diagram with the field galaxy 
distribution in the same space. This test will give the percentage of 
representativity
of the structure galaxy distribution compared to the field galaxy
distribution. The higher the percentage, the more different the
two distributions (value KS1 in Tables 4 and 5). We note that this test can be affected by inhomegeneous
galaxy sampling, inducing observational statistical differences between the
whole and the structure population.

\subsection{Structure galaxy content}

The galaxy content of the structures we detected is very important to
estimate because it allows us to understand how the galaxy structures are
evolving with redshift, mass and environment and how old these are. The key question is here
to estimate the galaxy type. We have used three independent methods
detailed in the following.

\subsubsection{Spectral characteristics}

First, we used the presence of
emission or absorption lines in galaxy spectra, given that pure
emission line galaxies are preferentially late type galaxies and that
pure absorption line galaxies are preferentially early type galaxies
(see e.g. Biviano et al. 2002). 

Using the VVDS spectra only (most of our sample), we were, therefore, able to
distinguish between galaxies with emission lines, absorption lines or both
(between z=0.2 and z=1.15: the redshift range where we
detected structures with VVDS data only). Then, we computed the percentage of
pure emission and absorption line galaxies in our catalog of
structures. We kept only structures with more than 5 spectra and at
least half of the structure population with a spectrum (as we have
only access to the VVDS spectra).

However, the detectability of a given emission or absorption line is a
complex interplay of several factors. Most of the time we are not able
to  follow a spectral line from z=0.2 to z=1.15 (due to our limited
spectral range: $\sim$[5500A;9400A]). For distant (and faint)
galaxies, emission lines are also usually easier to detect than absorption
lines. A simple representation of the variation with
redshift of the percentage of emission and absorption lines would,
therefore, be biased. For example, for z$\ge$0.9 galaxies, the only
major visible emission line is [OII]: this will artificially lower the number of
emission line-detected galaxies.
In order to remove these effects, we also computed the percentage of
pure emission and absorption line galaxies in our catalog of field
galaxies as a function of redshift. This allowed us to compare the structure galaxy content with
the field galaxy content in order to determine if,
for a given redshift, a structure galaxy content is morphologically earlier than
the field population.

\subsubsection{Photometric type}

Second, we used rest frame colors computed using Combo17 magnitudes
(Wolf et al. 2004 and see Ilbert et al. 2005). These colors computed in the AB
system using B and I
rest frame magnitudes are related to galaxy types according to Wolf et al. (2004). 
We divided these types into 2 bins: Elliptical + early spirals and late
spirals + Irregulars. We used only structures with more than 5
galaxies with an estimated morphology. 

\subsubsection{Morphological type}

Third, we used the classification made by Lauger et al. (2005)
for galaxies in the ESO/GOODS CDFS area
(Giavalisco et al. 2004).  This method is based on asymmetry and
concentration and a visual inspection of the galaxies from HST data in
several bands (rest-framed in the B band). We only review here the
salient points of this method:

        - the classification was made using the F435W, F606W, F775W and F850LP
HST filters in the B rest-frame passband

        - the method is based on the
Asymmetry (A) and light concentration (C) estimate (e.g. Bershady et al. 2000)
calibrated in the (A,C) parameter space by eye-balled morphological
types (using the same HST ACS data)

        - the basic method is only able to discriminate between bulge
          dominated (assumed to be early type) and disk dominated (assumed to be
          late type) objects. We adopted the same
          discrimination as in Lauger et al. (2005) and Ilbert et
          al. (2005) (see also Fig.~\ref{fig:morph}). 

We finally computed (using a Kolmogorov-Smirnov 2D test) the
probability of the A/C distributions of
structure galaxies in Fig.~\ref{fig:morph} to be different from the
field galaxy distribution in structures with more than 5 estimated
morphological types (see KS2 values in Tables 4 and 5). 

%\subsubsection{Comparison between morphological types}
%
%
%Fig.~\ref{fig:morph2} gives the excess percentage of early type
%galaxies compared to the field versus redshift for all the structures
%detected in the CDFS field of view. This shows the general good
%agreement between the 3 methods and gives confidence in our estimates
%of the early or late characteristics of the structure galaxies.

\subsection{Richness estimate}

When detecting a structure with the spectroscopic method, we produce a catalog
of the galaxies which are within the structure. However, it is not trivial to
discriminate between galaxies really belonging to this structure and just
passing-through galaxies (at the same structure redshift but not physicaly
bounded with the structure potential). This would require larger samples of
redshifts (of the order of 50 redshifts per structure: e.g. Mazure et
al. 1996). This is impossible with our data due to the high redshift of the
detected structures combined with the magnitude limit, the relatively low
sampling rate and the low-richness nature of the detected structures.

A possible resulting problem is the inclusion in the structures of non-member
galaxies. This can artificially increase the structure richness if estimated 
with the number of included galaxies (we are simply including field galaxies
inside the structure) or with the velocity dispersion (passing through
galaxies should have high relative velocities, increasing the velocity
dispersion estimate). We had therefore to find a way to compare the structure 
class (for a given redshift) estimated via the number of galaxies included or estimated via the
velocity dispersion. The number of spectroscopically 
measured galaxies in a given structure depends on:

       - the number of available targets down to the survey limiting magnitude
       (interplay between the structure richness, the survey limiting
       magnitude and the structure redshift)

       - the galaxy sampling rate in the survey inside the structure region
  
Using simulated cones (Rizzo et al. 2004) with richness-, redshift- and 
position-controled included clusters, we were able, using the VVDS SSPOC tool
(Bottini et al. 2005), to compute numbers of targeted galaxies down to I=24
theoretically present in our detected structures given their redshift, velocity
dispersion and position in the field (that give the sampling rate). Results
are shown in Table 3. This
allowed us to determine for the most interesting structures if the populations we
detected were reasonably rich. 

\section{Detected structures}

\subsection{The z$\sim$0.73 wall}

When applying  our spectroscopic structure detection algorithm, we tuned the
parameters to detect compact cluster-like structures. 
We detected with these parameters a wall at z=0.735 already detected
by Gilli et al. (2003), extending across the whole field 
of view  covered with spectroscopic redshifts (about 9 Mpc), with a velocity
dispersion as low as 665$\pm$116 km/s and sampled with 145 galaxies. This
structure is as compact in redshift as 
a cluster of galaxies and as extended on the sky as a supercluster of
galaxies. Such a structure could be also interpreted as a collapsing 
sphere of comoving radius of a few Mpc, producing a velocity dispersion close
to the one observed. In order to check this point, we used the whole field of view
with available photometric redshifts (not only spectroscopic redshifts). This
field is 1.9 larger in alpha and 1.4 times larger in delta, covering about
15Mpc. Fig.~\ref{fig:histoinout} shows that the z=0.73
structure is still very significantly visible outside of the spectroscopic
area. This leads to conclude that this structure is probably too large to be a simple
collapsing sphere and is really a wall.

We note that using the Rostat package (Beers et al. 1990), we detect 9 
significant empty redshift gaps inside the redshift 
distribution of the z=0.735 wall. This, added to the fact that there is
no extended 
X-ray emission across the whole field of view and that there is no visible red
sequence in the CMR (see Figs.~\ref{fig:cdfsinter} and
\ref{fig:cdfshigh}), shows that this
structure is 
not virialized. The galaxy velocity dispersion has, therefore, not to be
taken as a measure of the mass of the system. 

One region of the wall is particularly interesting as we clearly see
the presence of a more compact object in the center (see Fig.~\ref{fig:11-1fig}). 
This structure is 
also an extended X-ray source (XID 566 in Giacconi et al. 2002).

In order to study more carefully the structure of the wall, we applied
the Serna-Gerbal method (Serna $\&$ Gerbal 1996). This method allows
the detection of substructures in galaxy clusters, estimating the link
energy between galaxies of a compact structure. We detected 5
significant substructures inside Structure 11 (see Table 5). These structures
appear both in galaxy density
maps and galaxy color maps (for Structures 11-1, 11-2 and 11-4) computed using
photometric redshifts.

The largest one is identified with the central group
and associated with the extended X-ray source (XID 566). Sampled with
10 redshifts, this structure (Structure 11-1) has a velocity
dispersion of 455$\pm$161 km/s. This is typical of a low mass
cluster of galaxies (or a quite massive group). The Rostat package does not
detect any significant empty redshift gap in the redshift distribution of the galaxies
inside this structure. However, given the sampling rate at the
place where this structure is detected and the redshift of this structure, we
computed through simulations that such a $\sim$450km/s structure should be
sampled with 20-30 objects. This is an indication that the velocity dispersion
of structure 11-1 is probably overestimated. From the bolometric luminosity of
0.11 10$^{43}$ erg/s for this structure as computed from Giacconi et
al. (2003) data, we can derive an independent estimate of the mass. Following 
for example Jones et al. (2003), this is typical of
a normal group, with velocity dispersions around 200-300 km/s.

The galaxy content of this structure is essentially made of
early type galaxies from morphological estimates (see Fig.~\ref{fig:morph}),
very different from the field galaxy content, with a couple of central
galaxies probably in a merging process (fig 13 of Giacconi et
al. 2002). We used only these early type galaxies (from morphological
estimates) in  
Figs.~\ref{fig:cdfsinter} and \ref{fig:cdfshigh} to define the red sequence in
the CMR.

The red sequence in the CMR is well defined for Structure 11-1 (see
Figs.~\ref{fig:cdfsinter} and
\ref{fig:cdfshigh}) and the galaxy distribution in the
color/magnitude space is different from the field at the 99$\%$ level
using both V/V-R and R/R-I. The red sequence for Structure 11-1 has also the
correct position for z=0.735 (see Table 2).

All these arguments concur to show that we have detected a real low mass
relatively old structure lying in the core of the z=0.735 wall.

The 4 other substructures (Structures 11-2, 11-3, 11-4 and 11-5) are
sampled by 5 or 6 redshifts without extended X-ray counterparts. The
corrected velocity dispersions of these structures ranges from 150 to 500 km/s
but are typical of non virialized groups of galaxies given the
abscence of X-ray emission (see Table 5). These groups appear in 
Fig.\ref{fig:specphot2} to be preferentially constituted of red (and probably
early type) galaxies. This map has to be compared with Fig.\ref{fig:S11bis}
showing early type galaxies from the morphological classification. At least
for Structure 11-1, early type galaxies and red galaxies trace this same
central structure. Other structures inside the wall are not very well traced
by morphologicaly classified galaxy, but this is due to the fact that HST ACS
data are not covering the corner edges of the CDFS field.

\begin{figure*}
\centering
\mbox{\psfig{figure=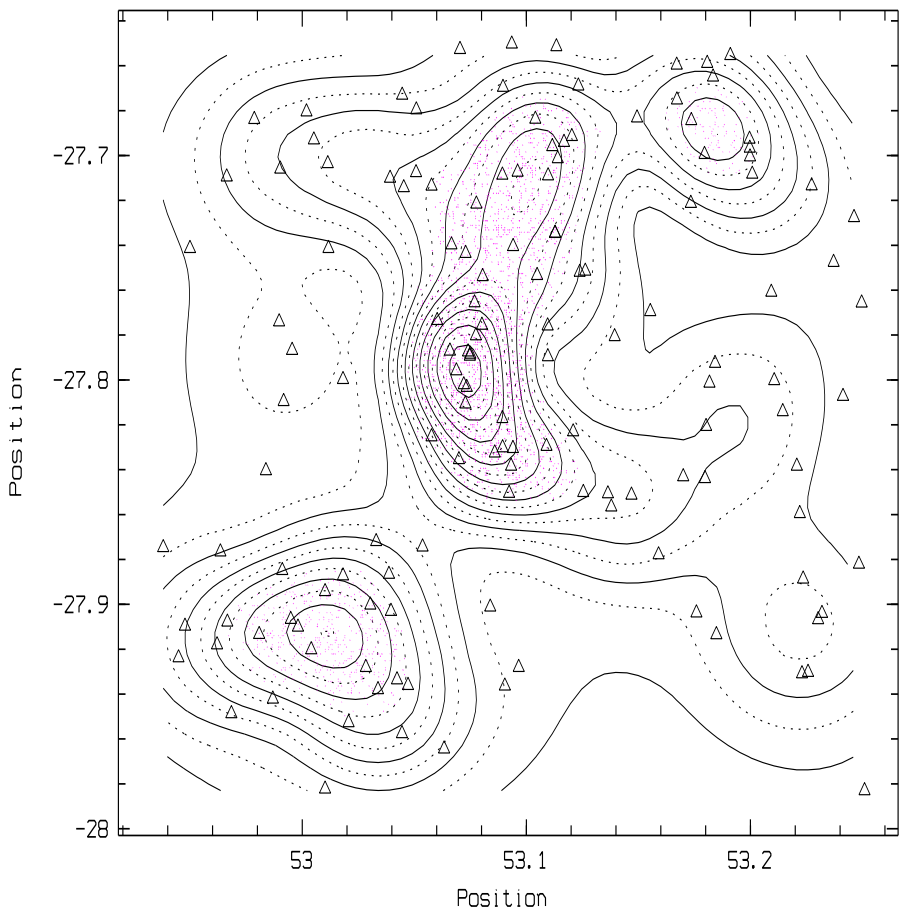,width=13.cm,height=13.cm,angle=270}}
\caption[]{Isocontours of the galaxy (with a measured redshift) density
across our CDFS field of view and inside the z=0.735 wall. Alpha and
Delta coordinates are given. Triangles are the individual galaxies
with a redshift inside the z=0.735 wall. Shaded areas are the areas
where the galaxy density estimate is significant at the 3-$\sigma$ level.}
\label{fig:mur2D}
\end{figure*}

We computed an adaptative kernel 2D map of the z$\sim$0.735 wall
galaxy density following the same recipe as 
described for example in Adami et al. (1998b) (see
Fig.\ref{fig:mur2D}) using first only galaxies with a
measured redshift. Triangles are the galaxies. Shaded areas are 
the places where the galaxy density estimate is significant at the
3-$\sigma$ level (from 1000 bootstrap resamplings), i.e. where the galaxy
overdensity is significant. This map is very similar to
Fig.\ref{fig:specphot2} (on a slightly larger area). All structures are
present in both maps, except for the West structures that are not visible with
spectroscopy due to low galaxy sampling rate. The central group is for example
clearly appearing in both maps (structure 11-1). Instead of 
a continuous wall, we detect using spectroscopy several other galaxy concentrations
roughly aligned from South West to North East. Is it due to inhomogeneous redshift 
sampling? The map computed with photometric redshifts (and therefore a $\sim$100$\%$
sampling rate) shows for
example a clear galaxy concentration (at coordinates -27.84, 52.96) where we 
did not detect any structure using spectroscopic redshifts. This was clearly 
due to low redshift sampling rate. This structure is not aligned with a  
South West - North East direction. 

The z=0.735 wall appears therefore as a central 
core (the detected central group) surrounded by a relatively isotropic large
accretion area. Such central
structures could be the progenitors of the most massive nearby clusters.

\subsection{A massive structure at z$\sim$1.10}

This structure at z=1.098 is a spatially compact structure
(the smallest in our sample) with an intermediate velocity 
dispersion of 373$\pm$131 km/s computed using 12
redshifts. This source is not detected as 
extended by Giacconi et al. (2002), but 2 sources classified 
as ponctual (but with possibly extended shapes) are clearly identified
with this structure (XID87 and 51). The XID51 source does not seem to have 
a thermal spectrum while the XID87 is possibly thermal according to its soft
and hard X-ray flux given in Giacconi et al. (2002). The bolometric luminosity
for XID87 assuming such a thermal spectrum is 0.37 10$^{43}$ erg/s. This is
typical of a massive group (e.g. Jones et al. 
2003) with an X-ray temperature of 1 or 2 keV. This is in good agreement with
the velocity dispersion estimate.

This structure (see Fig.~\ref{fig:S15spe}) is clearly bimodal and has a
complex redshift histogram. The Rostat package detects one significant gap in
the redshift histogram. Removing galaxies of the southern blob (difference of
0.0013 in redshift between the southern blob and the main structure) 
removes this gap but does not change significantly the velocity dispersion.
We should sample with 5 to 9
galaxies a $\sim$400 km/s structure at z$\sim$1.1 with the sampling rate we 
have. This is fully consistent with the really detected number of galaxies (8
galaxies if removing the southern blob).

Even if the CMR of Structure 15 does not show any evident red sequence, the
galaxy distribution in the
color/magnitude diagram is different from the field (at the 94$\%$ level).
Moreover, the brightest galaxy of the structure is classified as early
type using photometric classification. This galaxy has both emission
and absorption lines.

Regarding global galaxy content, the use of spectral features shows that we
have 2 times more pure absorption line galaxies compared to the field
and 2 times less pure emission line galaxies. Using photometric types
shows, however, a galaxy content close to the field. 

Using galaxy density maps (with photometric redshifts), we see a galaxy
overdensity close to the position of Structure 15. This overdensity also
appears in a color map (showing in an other way that galaxies around z=1.1 are
redder at this place).

We are probably seeing a future massive cluster in its formation stage, with already a
noticeable mass, typical of a massive group.

\subsection{Other detected compact structures}

The characteristics of other detected structures are listed in Table 4. These all
appear to be poor structures (and some of them could be fake detections). These
structures have velocity
dispersions ranging from 150 to 500 km/s and do not have associated X-ray
emission from Giacconi et al. (2002). The number of galaxies detected in these
structures is less than 10 except for Structure 2. This structure has a
velocity dispersion of 467$\pm$196 km/s (typical of a cluster of galaxies) 
without any significant gaps in the redshift histogram. However, there is no 
X-ray extended counterpart, raising doubts onto the velocity dispersion 
estimate, that is probably overestimated. This structure does not appear as
well proeminently in the color maps.

We note that all these structures are compact in redshift space (no gap detected by ROSTAT
(Beers et al., 1990)).

\subsection{Structure of the wall at z$\sim$0.66}

Already detected by Gilli et al. (2003), this wall (Structure 9) is much less compact
compared to the z=0.735 one and is embedding Structure 9-1.
To detect it as a whole, we had to relax the parameters of the
spectroscopic method to allow detection of larger and diffuse structures.
The structure is sampled with 66 redshifts and the 
global velocity dispersion is 1269$\pm$181 km/s with 10 significant empty redshift gaps
detected by the ROSTAT package (Beers et al. 1990). This structure has no
extended X-ray emission detected by Giacconi et al. (2002). This is,
therefore, not a virialized structure. 

A remarkable more compact structure (Structure 9-1) is detected inside this wall. This is
the only compact structure detected in this wall (Structure 10 being perhaps
an infalling group on the wall).
Well sampled with 11 redshifts, Structure 9-1 has a filamentary structure (see
Fig.\ref{fig:qsoX}) and a velocity dispersion of 
344$\pm$171 km/s. This would be typical of a group of galaxies, however, we
have no firm indication of a possible X-ray emission because this source is
just at the limit of the area covered by the Chandra data of Giacconi et
al. (2002). However, these data show an X-ray source
(J033219.5-275406, XID 249 in Giacconi et al. 2002) corresponding to a bright
galaxy close to Structure 9-1. This X-ray source has a hard spectrum typical
from an AGN and also
have a radio counterpart: NVSS J033219-275406. The optical counterpart object
is located at the end of the filament designed by Structure 9-1.
We do not have any spectroscopic redshift for this object, however, 
when computing a photometric redshift, the
best fit is obtained for a QSO template, also acting in favor of an "active" object. 
There is two solutions in redshift: 0.28 and 0.70 (see 
Fig.\ref{fig:qso20858}). The redshift
of this object is determined by an emission enhancement around
8250A. Associating H$\alpha$ with this emission will give z=0.28 and
associating H$\beta$ will give z=0.70 (in agreement with the Structure 9-1 redshift). 

The galaxies detected in Structure 9-1 both have emission and absorption
lines and are classified as late type galaxies using photometric and
morphological types. The CMR of Structure 9-1 appears also poorly
defined. This structure has 1.5 more emission
line galaxies than the field at the same redshift and the same number of absorption line
galaxies. The same tendency is visible using photometric and morphological
types. The galaxy content of
this structure is therefore quite late and similar to the field (using
morphological types, Structure 9-1 galaxy content is different from the
field galaxy content in the A/C diagram only at the 56$\%$ level).
Moreover, Structure 9-1 presents an interesting chain  
morphology very different from the axisymmetric usual cluster/group shape. 
Finally, this structure is not detectable in galaxy density or color maps (using
photometric redshifts).

The structure we have detected seems to be a peculiar case, quite
similar to the highly anisotropic compact and moderatly massive
structures proposed by West (1994) that are possibly the
progenitors of the giant nearby elliptical galaxies or of the nearby
fossil groups (e.g. Jones et al. 2003).

\begin{table*}
\caption{Number of galaxies in cluster and in the field (along the cluster line
  of sight) down to I$^{*}_{AB}$+2 for z=0.5
  and down to I$^{*} _{AB}$ for z=1
  for a Coma-like cluster and for several areas. We used I$^{*} _{AB}$$\sim$20.2
  at z=0.5 and I$^{*} _{AB}$$\sim$22.7 at z=1.}
\begin{tabular}{lllll}
\hline
Considered angular size & 0.8'$\times$0.8' & 2'$\times$2'  & 6'$\times$6' &
9.5'$\times$9.5'\\
Corresponding physical size at z=0.5 & 290kpc$\times$290kpc & 730kpc$\times$730kpc &
2200kpc$\times$2200kpc & 3480kpc$\times$3480kpc \\
Cluster/field ratio at z=0.5 & 11/5 & 36/32  & 108/288 & 127/720  \\
\hline
Considered angular size & 0.7'$\times$0.7' & 1.7'$\times$1.7'  & 5'$\times$5' &
7.5'$\times$7.5'\\
Corresponding physical size at z=1 & 340kpc$\times$340kpc & 820kpc$\times$820kpc &
2400kpc$\times$2400kpc & 3600kpc$\times$3600kpc \\
Cluster/field ratio at z=1 & 11/5 & 36/32  & 108/275 & 127/618  \\
\hline
\end{tabular}
\label{tab:key}
\end{table*}

\begin{table}
\caption{Elliptical galaxy colors for several redshifts. Magnitude bands are
  chosen to emcompass the 4000$\AA$ break given the redshift.}
\begin{tabular}{llll}
 & B-V & V-R & R-I \\
\hline
z=0.2 & 1.15 &  &  \\
z=0.4 & 2.20 & 1.25 &  \\
z=0.6 & & 1.16 & 1.31 \\
z=0.8 & & & 1.77 \\
z=1.0 & & & 1.81 \\
z=1.2 & & & 1.78 \\
\hline
\end{tabular}
\label{tab:key}
\end{table}

\begin{table}
\caption{Number of expected measured galaxies in clusters with various
  velocity dispersions as a function of the Global Sampling Rate (GSR). Values
  are given for z=0.73 and z=1.10.}
\begin{tabular}{lllll}
z & GSR & 400km/s & 600km/s & 900km/s \\
\hline
0.73 & 12$\%$ & 12 & 12 & 12 \\
0.73 & 23$\%$ & 20 & 23 & 23 \\
0.73 & 33$\%$ & 30 & 30 & 30 \\
0.73 & 43$\%$ & 35 & 43 & 48 \\
1.10 & 12$\%$ & 2 & 7 & 6 \\
1.10 & 23$\%$ & 5 & 11 & 10 \\
1.10 & 33$\%$ & 9 & 13 & 17 \\
1.10 & 43$\%$ & 16 & 15 & 22 \\
\hline
\end{tabular}
\label{tab:key}
\end{table}

\begin{table*}

\caption{Structures detected with coordinates (if not extended over the whole
  field of view), mean redshift, velocity
dispersion 
(km/s: vd) and its 1-$\sigma$ error, number of galaxies with a redshift belonging to the
structure, physical size (kpc), X-ray id from 
Giacconi et al. (2002), Kolmogorov Smirnov percentage (KS1) with the
considered color-magnitude space,  Kolmogorov Smirnov percentage (KS2) within the
A/C space (when more than 5 galaxies provided such a classification) and Class
following Section 5.}

\begin{tabular}{llllllllllll}

\hline

Id &alpha & delta &  z &   vd & Error
&    N & size  &   Xid Giacconi & KS1 & KS2 & Class \\

 & & &    & km/s
& km/s &    & kpc  &  & & &  \\

\hline

1 & 03 32 35.2 & -27 45 11.9   &   0.215    &  467 & 196 &   11   &   699
&      
-       & 99$\%$  B/B-V & 11$\%$ & 3 \\
2 & 03 32 14.5 & -27 48 39.2   &   0.227  &  265   & 89 &   5
&   658 &      
- & 41$\%$ B/B-V & 7$\%$& 4 \\
3 & 03 32 32.1 & -27 42 08.4    &  0.312   & 509  & 345 &  5   &   620
&     
-       & 97$\%$  B/B-V & & 3 \\

4 & 03 32 11.0 & -27 41 49.8   &   0.420   &  376  & 259 &  7   &   608
&     
-       & 93$\%$  B/B-V & & 3 \\

5 & 03 32 07.6 & -27 44 09.0   &   0.544   &  384  & 127 &  5
&   700 &     
-       & 64$\%$  V/V-R & 45$\%$ & 4 \\

6 & 03 32 01.2 & -27 45 19.2   &   0.576   &  312  & 97 &  5   &   677
&      
-       & 96$\%$  V/V-R & & 4 \\

7 & 03 32 29.1 & -27 55 55.9   &   0.619   & 340  & 148 &  9   &   1314
&     
- & 96$\%$  V/V-R & & 3 \\

8 & 03 31 55.4 & -27 41 51.9   &   0.621   &  398  & 125 &  5   &   702
&    
out of field & 70$\%$  V/V-R & & 4 \\

9 &  &   &   0.660   & 1269 & 181 &  66
& $\sim$8000  &     
- & 85$\%$  V/V-R & 9$\%$ & Wall \\

10 & 03 32 28.8 & -27 52 46.8   &   0.681   &  320  & 82 &   5
&  639 &    
- & 69$\%$ V/V-R & 12$\%$ & 3 \\

11 &  &  &   0.735   & 665   & 116 &  145
& $\sim$9000  &    
- & 99$\%$  R/R-I  & 6$\%$ & Wall \\

12 & 03 32 17.6 & -27 42 48.4   &   0.978   &  280  & 145 &   6
&   618 &    
- & 38$\%$  R/R-I  &  & 4 \\

13 & 03 32 29.7 & -27 43 01.0   &   1.036   &  329  & 289 &   7   &   1147
&     
- & 71$\%$ R/R-I & & 3 \\

14 & 03 32 16.6 & -27 51 51.9    &  1.044   &  154  & 84 &   7   
&   731 &     
- & 77$\%$ R/R-I & & 3 \\

15 & 03 32 14.7 & -27 52 58.7   &   1.098   &  373  & 131 &   12 &   440
&     
87 and 51 & 94$\%$ R/R-I & & 1 \\

16 & 03 32 19.8 & -27 46 23.52  &    1.221   &  161  & 149 &  6   &   791
&      
- & 53$\%$ R/R-I & & 3 \\

17 & 03 32 22.3 & -27 45 14.7   &   1.306   & 224 & 100 & 5   &   866
&      
- & 57$\%$ R/R-I & & 3 \\

\hline

\end{tabular}

\label{tab:col}

\end{table*}

\begin{table*}

\caption{Structures detected inside the z=0.66 and z=0.735 walls. Same label
  as Table 4.}

\begin{tabular}{llllllllllll}

\hline

Id &alpha & delta &  z &   vd & Error 
&      N & size &   Xid Giacconi & KS1 & KS2 & Class \\

& & &    & km/s & km/s 
&     & kpc &  & & & \\

\hline

9-1 & 03 32 32.8 & -27 59 08.5   &   0.660   &   344 & 171 &  11
&   723 & 641    
 & 64$\%$ V/V-R & 56$\%$ & 2 \\

11-1 & 03 32 20.7 & -27 46 05.7   &   0.736   &  455  & 161 &  10   &
941    &      566 & 99$\%$ V/V-R& 99$\%$ & 1 \\

11-2 & 03 31 55.9 & -27 54 35.2   &   0.736   & 496   & 132 &   5   &
659    &   - & 42$\%$ V/V-R& & 3 \\

11-3 & 03 32 27.5 & -27 41 18.0   &   0.734   &  157  & 53 &   6   &
376    &   - & 11$\%$ V/V-R & & 3 \\

11-4 & 03 32 51.8 & -27 42 55.0   &   0.732   &  315  & 63 &   6   &
800    &   - & 85$\%$ V/V-R& & 3 \\

11-5 & 03 32 54.4 & -27 54 22.6   &   0.736   &  399  & 143 &   5   &
800    &   - & 59$\%$ V/V-R& & 3 \\

\hline

\end{tabular}

\label{tab:col}

\end{table*}

\section{Discussion and Conclusions}

We detected 17 compact structures (Structure 11 being a compact wall split in 5 smaller
structures: see Table 5) and 1 more diffuse wall including one of the compact
structures. These structures are distributed all across
the CDFS field of view and have redshifts generally in good agreement with
the redshift peaks of the histogram of all galaxies along the
CDFS line of sight (Fig.~\ref{fig:adz}).

We detected a chain-like structure embedded in a quite diffuse wall at z=0.66
(structure 9-1) showing
signs of ongoing collapse and perhaps similar to the progenitors of giant
nearby elliptical galaxies.

We also have detected a dense wall at z$\sim$0.735, very compact in redshift
space and extending all across the field of view. The
existence of such extended and compact
structures in redshift space is remarquable as the
thickness of the structure in redshift space is very small 
for a non-virialized structure. 
However, as outlined for example by Kaiser (1987), such structures
have probably a strongly anisotropic clustering pattern. This results
in a compression of the structure along the line of sight, making 
these structures appear thinner in redshift space than
in real space (while in more massive and dense virialized structures as
clusters of galaxies, the effect in reversed, forming the well known
``fingers of god''). This structure is
interpreted as a central core (Structure 11-1) and an accretion area
composed of several bodies.

Among the structures we detected, we distinguish 4 classes: 

- 2 are real groups of galaxies (class 1): Structure 11-1 and 15

- 1 is a partially evolved and low mass structure of galaxies (class 2):
Structure 9

- 12 are proto-clusters/groups of galaxies (class 3): Structures 1, 3, 4, 7, 10, 11-2,
11-3, 11-4, 11-5, 13, 14, 16 and 17

- 5 are very early formation stage or fake structures of galaxies (class 4):
Structures 2, 5, 6, 8 and 12

Several arguments have already been discussed to check the ``reality'' of these
detected structures. In addition, we can discuss statistically
the redshift distribution of the identified structures. The more
relaxed a real structure will be, the more gaussian the structure
galaxy redshift distribution will appear. Due to small numbers inside
every individual group, we built synthetic distributions by
rescaling the redshift distributions of each structure: each redshift was
scaled using the mean redshift and the velocity dispersion of the
structure (see also Adami et al. 1998b). We
defined 3 subsamples: class 1 and 2 structures (50 galaxies), class 3
structures (67 galaxies) and class 4 structures (20 galaxies). First,
using Kolmogorov-Smirnov tests, we look for differences between the normalized
redshift distributions comparing each of them to the other two. 
Class 1+2 and class 3 sub-samples will only differ at
the 77$\%$ level (not very significant) while the class 4 sub-sample
differs at the 99$\%$ level from class 1+2 and from class
3 sub-samples. Moreover, fitting a gaussian to class 1+2 and class 3
sub-samples gives a reasonnable agreement, while it clearly fails for class 4
sub-sample. Therefore, class 4 sub-sample is possibly constituted of,
at least, partly of fake structures. 

The majority of the detected structures has a poorly defined red sequence in
the CMR, but significantly different from the field galaxy population.
In the hierarchical scenario, one would expect less evolved structures
with increasing redshift, with the distant ones more similar to
field galaxies in the color-magnitude diagram, but it has to be taken
into account the selection bias that richer structures are preferentially
detected with increasing redshift.

We also note that the structures detected at z$\ge$0.9 have a lower velocity
dispersion compared to the lower redshift sample: 254km/s versus
370km/s. Expressed in other terms the mean velocity dispersion of the
z$\le$0.9 structures is equal to the maximal velocity dispersion of the
z$\ge$0.9 structures. This would be consistent with the fact that we expect
less and less massive structures as the redshift is increasing (e.g. Evrard et
al. 2002). We also expect, however, the detection of more and more young
structures with increasing redshift, having, therefore, an artificially
increased velocity dispersion. This would imply an even stronger amplitude in our
observed velocity dispersion decrease with redshift.

Our sample field of view is by far too small in order
to efficiently constrain cosmological models using structure counts. However,
it is interesting to note that recent cosmological simulations (e.g. Evrard et
al. 2002) predict numbers for LCDM models in good agreement with our detections. Evrard et
al. (2002) predict for example between 1 and 4 structures more massive than 5
10$^{13}$ solar mass in our field of view (a $\tau$CDM model would predict
0 following Evrard et al.). Using for example Girardi et
al. (2002), such a mass is typical of Structure 15, being the only such one in
our sample at z greater than 1. Similarly, Evrard et al. (2002) predict no cluster at
all more massive than 3 10$^{14}$ solar mass in our field of view at z greater
than 1 and we detect no such clusters. Similar analyses will have,however, to
be performed on larger areas in order to give a reliable answer.

\begin{figure}
%\mbox{\psfig{figure=cl15bis.ps,width=8cm,angle=0}}
\caption[]{HST I image of Structure 15 with galaxies associated to the
structure (small circles) and the two X-ray sources (large circles). Size of
the image is 3.7'$\times$2.8'.}
\label{fig:S15spe}
\end{figure}

\begin{figure}
%\mbox{\psfig{figure=cl9bis.ps,width=8cm,angle=0}}
\caption[]{HST I image of Structure 9-1 with galaxies associated to the
structure (small circles) and the two X-ray sources (larges circles). Image
size is 5.5'$\times$3.3'.}
\label{fig:qsoX}
\end{figure}

\begin{figure}
\mbox{\psfig{figure=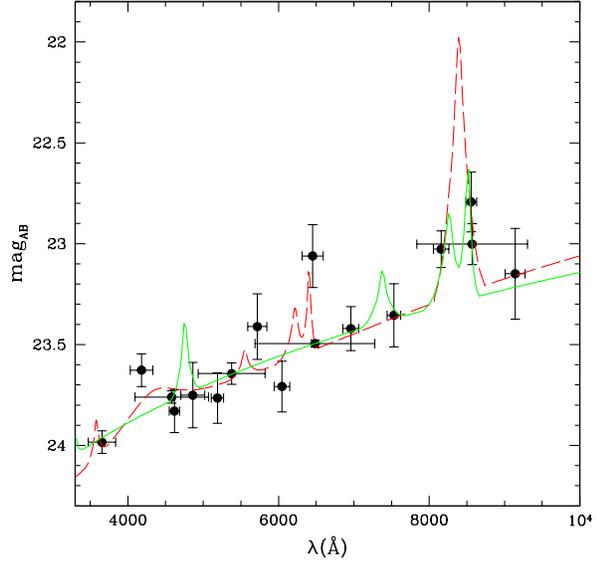,width=8cm,angle=0}}
\caption[]{Qso synthetic template fitted over Combo17 magnitudes for
  J033219.5-275406. Solid curve is the z=0.28 solution and dashed curve is the
  z=0.70 solution. }
\label{fig:qso20858}
\end{figure}

\begin{figure}
\centering
\mbox{\psfig{figure=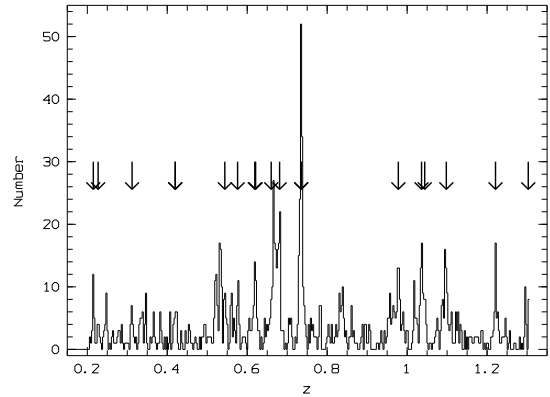,width=8cm,angle=270}}
\caption[]{Redshift histogram of the galaxies along
the CDFS line of sight. Redshift location of the detected
structures are shown as arrows. 
}
\label{fig:adz}
\end{figure}

\begin{acknowledgements}
This research has been developed within the framework of the VVDS
consortium.\\
This work has been partially supported by the
CNRS-INSU and its Programme National de Cosmologie (France),
and by Italian Ministry (MIUR) grants
COFIN2000 (MM02037133) and COFIN2003 (num.2003020150).\\
The VLT-VIMOS observations have been carried out on guaranteed
time (GTO) allocated by the European Southern Observatory (ESO)
to the VIRMOS consortium, under a contractual agreement between the
Centre National de la Recherche Scientifique of France, heading
a consortium of French and Italian institutes, and ESO,
to design, manufacture and test the VIMOS instrument.\\
C.A. thanks G. Lima-Neto for useful comments.\\
Authors thank the referee for useful and constructive comments.\\
\end{acknowledgements}


\begin{thebibliography}{}

\bibitem[]{} Abell G.O., Corwin H.G., Olowin R.P., 1989, ApJS 70, 1

\bibitem[]{} Adami C., Mazure A., 2002, A$\&$A 381, 420

\bibitem[]{} Adami C., Mazure A., Biviano A., Katgert P., Rhee G.,
1998, A$\&$A 331, 493 (a)

\bibitem[]{} Adami C., Biviano A., Mazure A.,
1998, A$\&$A 331, 439 (b)

\bibitem[]{} Arnouts S., Vandame B., Benoist C., et al.,
2001, A$\&$A 379, 740

\bibitem[]{} Baum W., 1959, PASP 71, 106

\bibitem[]{} Beers T., Flyn K., Gebhardt K., 1990, AJ 100, 32

\bibitem[]{} Bershady M.A., Jangren A., Conselice C., 2000, AJ 119, 2645

\bibitem[]{} Biviano A., Durret F., Gerbal D., et al., 1996, A$\&$A 311, 95

\bibitem[]{} Biviano A., Katgert P., Thomas T., 2002, A$\&$A 387, 8

\bibitem[]{} Bottini D., Garilli B., Maccagni D., et al., 2005, A$\&$A in press

\bibitem[]{} Castander F.J., 1998, ApSS 263, 91

\bibitem[]{} Cimatti A., Daddi E., Mignoli M., et al., 2002, A$\&$A 381, L68

\bibitem[]{} Dressler A., Smail I., Poggianti B., 1999, ApJS 110, 213

\bibitem[]{} Dav\'e R., Hellinger D., Primack J., Nolthenius R.,
Klypin A., 1997, MNRAS 284, 607

\bibitem[]{} Evrard A.E., MacFarland T.J., Couchman H.M.P., et al., 2002, ApJ 573, 7

\bibitem[]{} Folkes S., Ronen S., Price I., et al., 1999, MNRAS 308, 459

\bibitem[]{} Giacconi R., Zirm A., Wang J.X., et al., 2002, ApJSS 139, 369

\bibitem[]{} Giavalisco M., Dickinson M., Ferguson H.C., et al., 2004, ApJ 600, L93

\bibitem[]{} Gilli R., Cimatti A., Daddi E., et al., 2003, ApJ 592, 721

\bibitem[]{} Girardi M., Manzato P., Mezzetti M., et al., 2002, ApJ 569, 720

\bibitem[]{} Ilbert O., et al., 2005, A$\&$A  in preparation

\bibitem[]{} Jones L.R., Ponman T.J., Horton A., et al., 2003, MNRAS 343, 627

\bibitem[]{} Kaiser N. 1987, MNRAS 227, 1

\bibitem[]{} Lauger S., et al., 2005, A$\&$A  in preparation

\bibitem[]{} Lax D. 1985, J. Am. Stat. Assoc. 80, 736

\bibitem[]{} Le F\`evre O., Vettolani G., Paltani S., et al., 2004, A$\&$A 428, 1043

\bibitem[]{} Mazure A., Katgert P., den Hartog R., et al., 1996, A$\&$A 310, 31

\bibitem[]{} Moy E., Barmby P., Rigopoulou D., et al., 2003, A$\&$A
403, 493

\bibitem[]{} Rizzo D., Adami C., Bardelli S., et al., 2004, A$\&$A 413, 453

\bibitem[]{} Romer A.K., Viana P.T.P.., Liddle A.R., Mann R.G., 2001,
ApJ 547, 594

\bibitem[]{} Sandage A., 1972, ApJ 176, 21 

\bibitem[]{} Scaramella R., Bottini D., Garilli B., et al., 2006, A$\&$A in preparation

\bibitem[]{} Serna A., Gerbal D., 1996, A$\&$A 309, 65 

\bibitem[]{} Szokoly G.P., Bergeron J., Hasinger G., et al., 2004,
ApJS 155, 271

\bibitem[]{} West M.J., 1994, MNRAS 268, 79

\bibitem[]{} Wolf C., Meisenheiner K., Kleinheinrich M., et al. 2004,
A$\&$A, submitted astro-ph: 0403666


\end{thebibliography}
\end{document}